\documentclass[twocolumn, superscriptaddress, nofootinbib]{revtex4-1}

\usepackage[english]{babel}
\usepackage[pdftex]{graphicx}
\usepackage{color}

\usepackage[sort&compress]{natbib}
\usepackage{hyperref}
\usepackage{amsmath}
\usepackage{amssymb}
\usepackage{mathtools}
\usepackage{braket}

\renewcommand{\vec}[1]{\ensuremath{\boldsymbol{#1}}}
\newcommand{\un}[1]{\ensuremath{\,\mathrm{#1}}}
\newcommand{\fig}[1]{Figure~\ref{fig:#1}}
\newcommand{\eq}[1]{(\ref{#1})}

\newcommand{\I}{\mathrm{i}}
\newcommand{\abs}[1]{\left| #1 \right|}
\newcommand{\lr}[1]{\ensuremath{\left( #1 \right)}}
\renewcommand{\deg}{^{\circ}}
\renewcommand{\ap}{\alpha}
\renewcommand{\th}{\theta}

\begin{document}

\title{Electron optics in phosphorene pn junctions:\\Negative reflection and anti super-Klein tunneling}

\author{Yonatan~Betancur-Ocampo}
\email{ybetancur@icf.unam.mx}
\affiliation{Instituto de Ciencias F\'isicas, Universidad Nacional Aut\'onoma de M\'exico, Cuernavaca, M\'exico}

\author{Fran\c{c}ois~Leyvraz}
\affiliation{Instituto de Ciencias F\'isicas, Universidad Nacional Aut\'onoma de M\'exico, Cuernavaca, M\'exico}
\affiliation{Centro Internacional de Ciencias, Cuernavaca, M\'exico}

\author{Thomas~Stegmann}
\email{stegmann@icf.unam.mx}
\affiliation{Instituto de Ciencias F\'isicas, Universidad Nacional Aut\'onoma de M\'exico, Cuernavaca, M\'exico}

\begin{abstract}  
  Ballistic electrons in phosphorene $pn$ junctions show optical-like phenomena. Phosphorene is
  modeled by a tight-binding Hamiltonian that describes its electronic structure at low energies,
  where the electrons behave in the zigzag direction as massive Dirac fermions and in the orthogonal
  armchair direction as Schr\"odinger electrons. Applying the continuum approximation, we derive the
  electron optics laws in phosphorene $pn$ junctions, which show very particular and unusual
  properties. Due to the anisotropy of the electronic structure, these laws depend strongly on the
  orientation of the junction with respect to the sublattice. Negative and anomalous reflection are
  observed for tilted junctions, while the typical specular reflection is found only, if the
  junction is parallel to the zigzag or armchair edges. Moreover, omni-directional total reflection,
  called anti-super Klein tunneling, is observed if the junction is parallel to the armchair
  edge. Applying the nonequilibrium Green's function method on the tight-binding model, we calculate
  numerically the current flow. The good agreement of both approaches confirms the atypical
  transport properties, which can be used in nano-devices to collimate and filter the electron flow,
  or to switch its direction.
\end{abstract}

% Keywords: Phosphorene, pn junctions, electron optics, Klein tunneling, negative reflection

\maketitle

\section{Introduction}
\label{sec:intro}

Electron optics originates from the wave-particle duality and has lead to outstanding technological
applications like the electron microscope \cite{Pozzi}. The rise of graphene and other
two-dimensional materials has given electron optics recently a new turn, because ballistic ray-like
electron propagation can be observed in these materials at low energies. The electron rays can be
manipulated by external gates that form a $pn$ junction \cite{Cheianov, Williams, Rickhaus, Liu}. In
graphene, the electrons are refracted at the interface of the junction following a generalized
Snell's law, where the gate voltages in the different regions play the role of the refractive
indices and cause negative refraction \cite{Cheianov}. Moreover, perfect transmission at normal
incidence can be observed, which is known as Klein tunneling \cite{Katsnelson}. These phenomena have
been confirmed experimentally \cite{Kim, Lee, Chen}. Another recently discussed possibility to
manipulate the electron flow in graphene is to deform the material elastically \cite{Stegmann,
  Stegmann2, Betancur, Naumis, Naumis2, Carrillo, Andrade}. These discoveries made it possible to
propose new nano-devices such as superlenses \cite{Cheianov, Betancur, Betancur3}, valley beam
splitters \cite{Stegmann, Stegmann2, Pomar, Zhai, Zhai2, Peeters, Charlier, Settnes, Sa, Zhai3,
  Carrillo2} and collimators \cite{Stegmann, Liu, Park}.

Recently, a monolayer of black phosphorous called phosphorene has been synthesized \cite{Li}. Also
phosphorene nanoribbons have been produced \cite{Watts}. The electronic structure of this material
shows (in contrast to graphene) an intrinsic band-gap and strong anisotropy, which causes the
electrons to behave in one direction like massive Dirac fermions and in the orthogonal direction
like non-relativistic Schr\"odinger electrons \cite{Li2, Carvalho, Kou, Ehlen, Peng, Sisakht,
  Peeters2, Ezawa, Lewenkopf, Lew, Brown, Rudenko, Rudenko2, Cakir, Elahi, Partoens}. This unique
electronic structure makes phosphorene very attractive for both, fundamental research and
technological applications such as transistors \cite{Zhu, Koenig, Ong, Sarkar, Miao, Lv, Utt, Naemi}
and sensors \cite{Ray}.

In this article, we study theoretically the current flow in phosphorene $pn$ junctions and answer
the question how electron beams are reflected and transmitted at the interface of the junction. We
start from a simple second nearest neighbor tight-binding model that reproduces well the electronic
structure of phosphorene at low energies. Using this Hamiltonian, on the one hand, we calculate
numerically the current flow by means of the nonequilibrium Green's function (NEGF) method. On the
other hand, applying the continuum approximation, we calculate the semiclassical trajectories of the
electrons and derive the laws of reflection and refraction for electron optics in phosphorene $pn$
junctions. Both approaches agree qualitatively and demonstrate unusual optical phenomena like
\textit{negative and anomalous reflection}, which have been observed previously only for
electromagnetic waves at the surface of meta materials \cite{Liu2,Liu3,Yu,Yu2}. For certain system
parameters we also report omni-directional perfect reflection, called \textit{anti super-Klein
  tunneling}, which is the counterpart of the omnidirectional perfect transmission of massless Dirac
fermions in Dirac materials \cite{Betancur3} and pseudo-spin one systems \cite{Bercioux2, Shen,
  Betancur2, Fang}. Our theoretically work will pave the way to electron optics experiments in
phosphorene, which to the best of our knowledge have not yet been performed.

\section{Modeling phosphorene}
\label{sec:model}

\begin{figure*}[t]
  \centering
  \includegraphics[scale= 0.43]{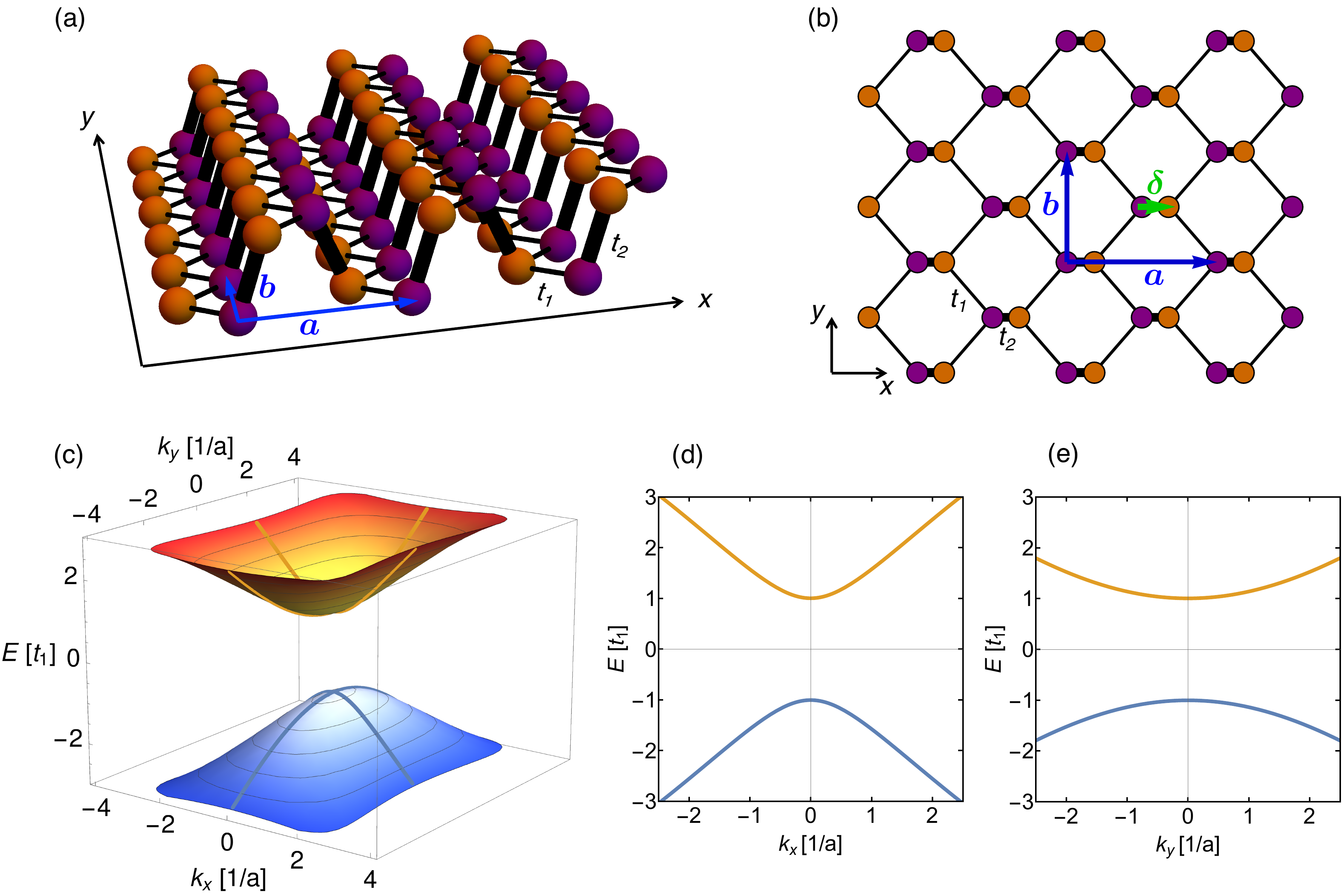}
  \caption{(a) Crystal structure of phosphorene. Its electronic structure is modeled by a
    tight-binding Hamiltonian that takes into account first ($t_1$) and second ($t_2 \approx -3t_1$)
    nearest neighbors. The vectors $\vec{a}$ and $\vec{b}$ span the part of the unit cell that is in
    the $xy$ plane. (b) Projection of all atoms to the $xy$ plane shows that phosphorene can be also
    understood as a deformed trigonal lattice with a diatomic basis, as indicated by the different
    coloring of the atoms. The vector $\vec{\delta}$ connects these two atoms of the unit cell. (c)
    The electronic band structure of phosphorene shows a band gap at the $\Gamma$ point and is
    highly anisotropic. The approximately linear dispersion of massive Dirac fermions is observed in
    the $k_x$ direction (d), while the parabolic dispersion of Schr\"odinger electrons is found
    along the $k_y$ direction (e).}
  \label{fig:1}
\end{figure*}

Phosphorene is a two-dimensional puckered crystal with four phosphorous atoms in the unit cell, see
\fig{1}~(a). A detailed introduction and an overview can be found in
Refs. \cite{Carvalho,Sisakht,Peeters2,Ezawa,Lewenkopf,Lew}. The electronic structure of phosphorene
is modeled by the tight-binding Hamiltonian
\begin{equation}
  H = \sum_{\braket{i,j}} t_{ij}\ket{i}\bra{j} + \textrm{H.c.},
\label{Hc}
\end{equation}
where the $\ket{i}$ represent the atomic states localized on the phosphorous atoms at positions
$\vec{r}_i$. The sum includes first and second nearest neighbors which are coupled by the energies
$t_1= -1.22 \un{eV}$ and $t_2= 3.665 \un{eV} $, respectively \cite{Rudenko,Rudenko2}.

The physical description of phosphorene can be simplified by projecting all atoms on the $xy$ plane
keeping the coupling energies constant \cite{Sisakht, Peeters, Ezawa}. After this projection, see
\fig{1}~(b), the phosphorene lattice can be understood as a strongly deformed honeycomb lattice,
where the armchair edge runs along the $x$ direction while the zigzag boundary is parallel to the
$y$ direction. This lattice is formed by trigonal basis vectors and has a unit cell with only two
atoms that are connected by the vector $\vec{\delta}$ of length $0.8\un{\AA}$. From this point of
view phosphorene is very similar to deformed graphene. However, the clear difference between the
hopping parameters ($t_2 \approx -3t_1$) causes very different physical properties. For simplicity
we will continue using the orthogonal basis vectors of the original phosphorene lattice, $\vec{a}$
and $\vec{b}$, which have a length of $4.42 \un{\AA}$ and $3.27 \un{\AA}$, respectively
\cite{Brown}.

\begin{figure*}[t]
  \centering
  \includegraphics[scale= 0.44]{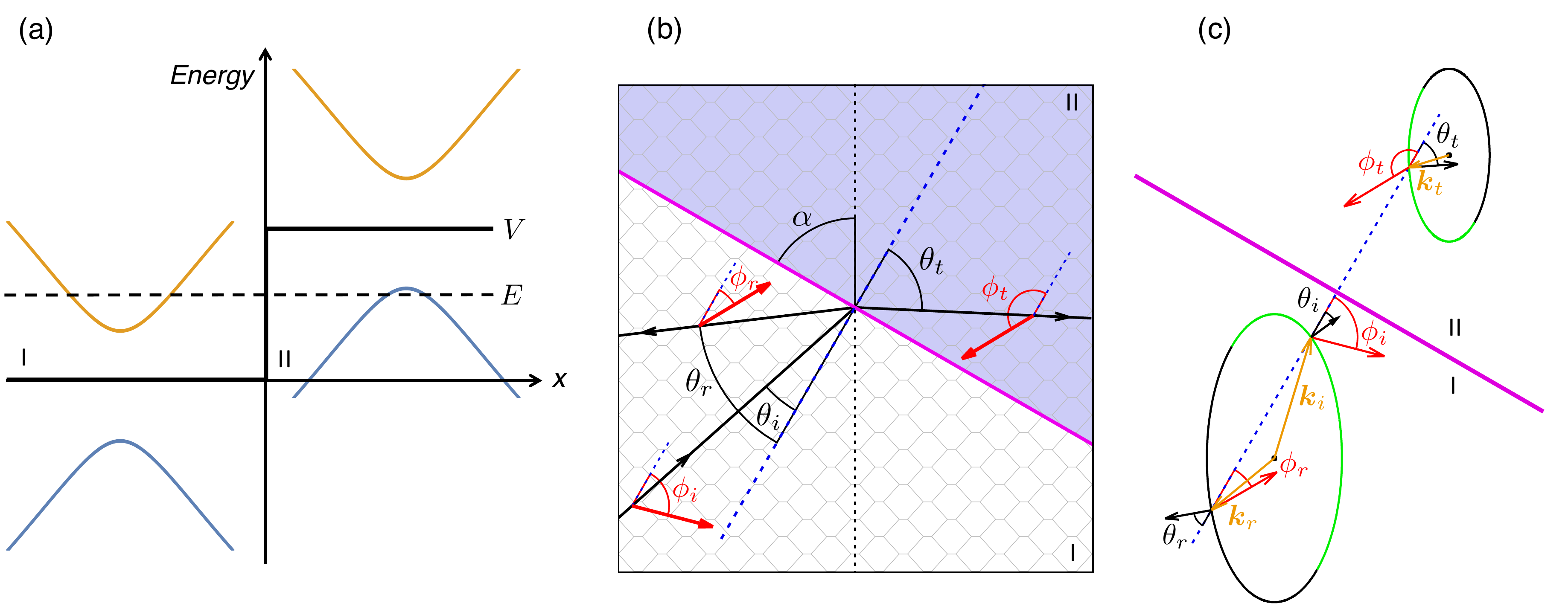}
  \caption{Electron optics in a phosphorene $pn$ junction.  (a) The gate-induced electrostatic
    potential $V$ shifts the energy bands of phosphorene and generates a $pn$ junction. In the left
    region I the electrons of energy $E$ belong to the conduction band (orange), while in the right
    region II they are in the valance band (blue). (b) The interface of the junction (violet solid
    line) is tilted by the angle $\ap$, which is an important parameter due to the anisotropy of
    phosphorene's electronic structure.  An incident electron beam (black solid line) with
    pseudo-spin $\phi_i$ (red arrow) hits the interface under the angle $\th_i$. It is reflected
    (atypically) under the angle $\th_r$ with pseudo-spin $\phi_r$, while it is transmitted under
    the angle $\th_t$ with pseudo-spin $\phi_t$. (c) Kinematical construction in the momentum
    space. The closed curves are constant energy contours. The current density (or group velocity)
    vectors are represented by the black arrows, which are orthogonal to the energy contours. The
    green semi-arcs indicate the regions where the current density $j_\perp$ is conserved. The
    blue-dashed line represents the conservation of the parallel momentum $k_{||}$. Its intersection
    points with the constant energy contours determine the reflected and transmitted states (orange
    arrows).}
\label{fig:2}
\end{figure*}

Making a plane wave ansatz, the tight-binding Hamiltonian can be written as
\begin{equation}
  H(\vec{k}) = \left(\begin{array}{@{}*{2}{c}@{}}
0 & g^*(\vec{k})\\
g(\vec{k}) & 0
\end{array}\right),
\label{H}
\end{equation}
where
\begin{equation}
  \label{gofk}
  g(\vec{k}) = \textrm{e}^{-\I k_{\delta}}[t_2 + 2t_1\textrm{e}^{\I k_a/2}\cos(k_b/2)]
\end{equation}
with $k_{\delta} = \vec{k}\cdot\vec{\delta}$, $k_{a} = \vec{k}\cdot\vec{a}$ and
$k_{b} = \vec{k}\cdot\vec{b}$. The energy bands of phosphorene are given by
\begin{equation}
  \label{eofk}
  \begin{array}{c}
    E(\vec{k})= s \abs{g(\vec{k})} \\[2mm]
    = s \sqrt{t_2^2 + 4 t_1 t_2 \cos(k_a/2)\cos(k_b/2) + 4 t_1^2 \cos^2(k_b/2)}
  \end{array}
\end{equation}
where $s = \text{sgn}(E)$ is the band index. The energy bands, see \fig{1}~(c)-(e), show a direct
band gap of $2 \Delta = 4t_1 +2t_2$ in the center of the Brillouin zone ($\Gamma$ point) and are
highly anisotropic. The corresponding eigenfunctions read
\begin{equation}
  \Psi_{\vec{k}}(\vec{r}) = \frac{1}{\sqrt{2}}\left(
    \begin{array}{@{}*{1}{c}@{}}
      1 \\
      s\textrm{e}^{\I \phi(\vec{k})}
    \end{array}\right)
  \textrm{e}^{\I \vec{k}\cdot \vec{r}},
  \label{wf}
\end{equation}
with the phase (or pseudo-spin)
\begin{eqnarray}
  \phi(\vec{k}) & = & \arctan\left(\frac{\textrm{Im}[g(\vec{k})]}{\textrm{Re}[g(\vec{k})]}\right) \nonumber\\[2mm]
       & = &\arctan\left(\frac{\lambda_1(\vec{k})\cos(k_{\delta}) -\lambda_2(\vec{k})\sin(k_{\delta})}
           {\lambda_2(\vec{k})\cos(k_{\delta}) +\lambda_1(\vec{k})\sin(k_{\delta})}\right),
\label{phis}
\end{eqnarray}
which is a function of $\lambda_1(\vec{k}) = 2t_1\sin(k_a/2)\cos(k_b/2)$ and
$\lambda_2(\vec{k}) = t_2 + 2t_1\cos (k_a/2)\cos (k_b/2)$. Note that the prefactor
$\textrm{e}^{-\I k_{\delta}}$ in \eq{gofk} is frequently omitted because it cancels out in the
calculation of the energy bands. However, the electron propagation will be described incorrectly if
this term is not included in the eigenfunctions, because their phase depends explicitly on
$k_{\delta}$.

The anisotropy of the energy bands of phosphorene is observed more clearly, if the Hamiltonian
\eq{H} is expanded around the $\Gamma$ point%
\footnote{Note that in contrast to graphene, where the Dirac cones are located at the six $K$
  points, the band gap in phosphorene appears at the single $\Gamma$ point in the center of the
  Brillouin zone.}
\begin{equation}
  H_{\textrm{eff}} = \sigma_x\left(\Delta +\frac{p^2_x}{2m_x} +\frac{p^2_y}{2m_y}\right) +\sigma_yvp_x,
  \label{Heff}
\end{equation}
where $ m_x= 2/\left( -t_1a^2 + 2\delta(2at_1 - \delta\Delta)\right)$ and
$m_y= -2/\left(t_1b^2\right)$ are the anisotropic masses, and $v = at_1 - \delta\Delta$ is the
velocity along the $x$ direction. The energy bands take the form
\begin{equation}
 E_s = s\sqrt{\left(\Delta + \frac{p^2_x}{2m_x} + \frac{p^2_y}{2m_y}\right)^2 + v^2p^2_x}
 \label{dr}
\end{equation}
and show clearly the hybrid behavior of the electrons in the different directions. If $p_y = 0$, the
electrons behave approximately as massive Dirac fermions with velocity $v_F = \sqrt{v^2+\Delta/m_x}$
and rest mass $m_0 = \Delta/v^2_F$. If $p_x=0$, the electrons have the parabolic dispersion of
Schr\"odinger electrons.

To conclude this section, the electronic band structure in \fig{1}~(c)-(e) shows that the used
second nearest neighbor tight-binding model reproduces well the essential electronic properties of
phosphorene at low energies, namely a band gap and strong anisotropy. Taking into account higher
orders modify the size of the band gap, but will not change qualitatively the shape of the
electronic structure (at low energies), which is the basis for electron optics to be discussed in
the following sections.

\section{Electron optics in phosphorene pn junctions}
\label{sec:eopt}

A phosphorene $pn$ junction is constituted by a monolayer of black phosphorous with two differently
doped regions, I and II, respectively. These regions, which are indicated in \fig{2}~(b) by
different background color shadings, are realized experimentally by gates that induce different
electrostatic potentials. These potentials shift the electronic structure and hence, change the
electron densities, see \fig{2}~(a). Without loss of generality, we assume that the electrostatic
potential is zero in the white-shaded region I, where the electrons will be injected, while it has
the value $V$ in the blue-shaded region II. At the interface between the regions the electrons are
reflected and refracted in different directions, similar to a light beam that passes from one medium
to another. Using the continuum approximation of phosphorene, we consider the interface as a
straight line without any corrugation due to the lattice structure. Moreover, we assume that at the
interface the potential changes abruptly on the length scale defined by the Fermi wavelength
$\lambda_F= 2\pi/|\vec{k}|$. This assumption is made because the band gap of phosphorene damps
drastically the transmission through a junction with a smooth potential profile.

As phosphorene is a material with an anisotropic electronic structure, the transport properties
through the $pn$ junction depend on the orientation of the junction with respect to the phosphorene
lattice, which is measured in the following by the angle $\ap$. The incident electron beam with
energy $E$ hits the interface of the junction under the angle $\th_i$. These four parameters, $V$,
$\ap$, $E$, $\th_{i}$, which have a clear physical meaning, will be used in the following to
characterize the system. All other quantities will be calculated as a function of these
parameters. For example, the wave vector $\vec{k}_i$ of the incident electrons is calculated by
using \eq{eofk} to obtain
\begin{equation}
  k_a = 2\arccos\lr{\frac{E^2 - t^2_2 - 4t^2_1\cos^2(k_b/2)}{4t_1t_2\cos(k_b/2)}}
  \label{ka}
\end{equation}
and by taking into account that the angle of incidence is related to the group velocity
$\partial E/ \partial \vec{k}$ by means of%
\footnote{We do not use here the expansion of the band structure around the $\Gamma$, because this
  causes already at relatively low energies discrepancies to the numerical simulations.}
\begin{equation}
  \frac{v_y}{v_x}= \tan\th_i = \frac{b}{a}\lr{\frac{\tan(k_b/2)}{\tan(k_a/2)}+\frac{2t_1\sin(k_b/2)}{t_2\sin(k_a/2)}}.
  \label{ths}
\end{equation}
These two equations allow to determine the two components of $\vec{k}_i$, as well as the pseudo-spin
$\phi_i$ by means of \eq{phis}.

The reflection and refraction laws of the electrons at the interface of the $pn$ junction are
determined by the conservation of the electron energy $E$, the parallel component of the momentum
$k_{||} = -k_x\sin\ap + k_y\cos\ap$ and the normal component of the current density $j_\perp$ (due
to the translational symmetry along interface and the continuity equation). These quantities are
sketched in the kinematical construction in \fig{2}~(c). The closed curves are constant energy
contours in the regions I and II. They have different size due to the electrostatic potential in
region II. The straight blue-dashed line is given by
$k_y = \left(k_x-k_{i,x}\right)\tan\ap + k_{i,y}$ and represents the conservation of $k_{||}$. Its
intersection points with the constant energy contours determine the wave vectors $\vec{k}_{r/t}$ of
the reflected and transmitted electrons. Note that the green semi-arcs indicate the regions where
$j_\perp$ is conserved and hence, the correct solution for $\vec{k}_t$. With these wave vectors, we
can calculate the pseudo-spins $\phi_r$ and $\phi_t$ from \eq{phis}, as well as the angles of
reflection $\th_r$ and refraction $\th_t$ using \eq{ths}, which constitute the electron optics laws
in phosphorene $pn$ junctions. An important consequence of the anisotropy of phosphorene's band
structure is the fact that the current density (group velocity), momentum, and pseudo-spin are not
always parallel. Therefore, the pseudo-spin angles $\phi$ and $\phi_r$ can not be interpreted as the
angles of incidence and refraction respectively, as is usually the case for isotropic systems like
graphene \cite{Fuchs}. This causes new and interesting phenomena to be discussed in
Section~\ref{sec:results}, which can be understood mainly by the fact that the pseudo-spin is
conserved in the $pn$ junction as the electrostatic potential does not break the sublattice
symmetry.

In order to obtain the probabilities of reflections $R$ and transmission $T$, we consider the wave
function in region~I
\begin{equation}
  \Psi_{\text{I}}(\vec{r}) = \frac{1}{\sqrt{2}}\left(
    \begin{array}{c}
      1\\
      s\textrm{e}^{\I \phi_i}
    \end{array}\right)
  \textrm{e}^{\I \vec{k}_i\cdot\vec{r}} +
  \frac{r}{\sqrt{2}}\left(
    \begin{array}{c}
      1\\
      s\textrm{e}^{\I \phi_r}
    \end{array}\right)
  \textrm{e}^{\I \vec{k}_r\cdot\vec{r}}
\label{pwr1}
\end{equation} 
and in region II
\begin{equation}
  \Psi_{\text{II}}(\vec{r}) = \frac{t}{\sqrt{2}}\left(
    \begin{array}{c}
      1\\
      s'\textrm{e}^{\I \phi_t}
    \end{array}\right)\textrm{e}^{\I \vec{k}_t\cdot\vec{r}},
\label{pwr2}
\end{equation}
where $s= \text{sgn}(E)$ and $s'=\text{sgn}(E-V)$ are the band indices in the two regions. The
amplitudes of reflection $r$ and transmission $t$ are calculated by using the continuity of the wave
function at the interface. Finally, due to the conservation of the probability $T=1-\abs{r}^2$, we
obtain for the transmission probability
\begin{equation}
  T = \frac{2\sin[\phi_t -(\phi_r+\phi_i)/2]\sin[(\phi_i-\phi_r)/2]}{ss'-\cos(\phi_t-\phi_r)}.
\label{T}
\end{equation}
Note that the transmission is given as a function of the pseudo-spins, which in turn depend on the
parameters $V$, $\ap$, $E$ and $\th$, see \eq{phis}, \eq{ka} and \eq{ths}. Hence, $T$ is an
(implicit) function of these four parameters as we will demonstrate in Section~\ref{sec:results}.

\section{The NEGF method for the current flow}
\label{sec:NEGF}

\begin{figure*}[t]
  \centering
  \includegraphics[scale= 0.45]{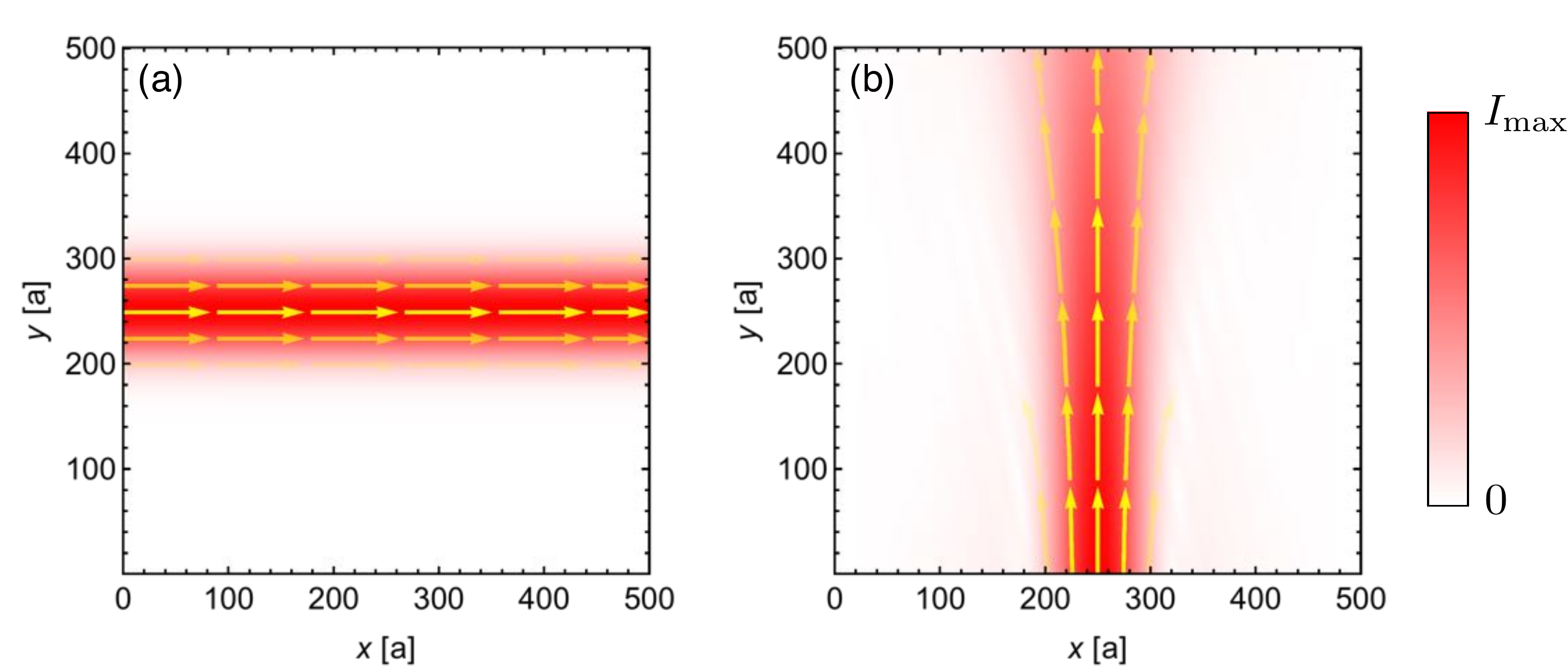}
  \caption{Current flow in a phosphorene nanoribbon without an electrostatic potential
    ($V=0$). Electrons with energy $E=1.2\abs{t_1}$ are injected at the left (a) and bottom (b) edge of
    the nanoribbon. The current density is indicated by the red color shading and its vector field
    by the yellow arrows. Ray-like ballistic electron propagation can be observed in both cases,
    although the diffraction of the electron beam is stronger in the $y$ direction (b) than in the
    $x$ direction (a).}
\label{fig:3}
\end{figure*}

Apart from the laws of reflection and refraction derived in the previous section, we will also
calculate numerically the quantum coherent current flow in the phosphorene $pn$ junctions, starting
from the tight-binding Hamiltonian of a finite system and applying the nonequilibrium Green's
function method (NEGF). As this method does not make use of the continuum approximation, it allows
us to confirm numerically the electron optics laws. Moreover, wave effects like diffraction and
interference are included which go beyond the ray-like trajectories of geometric optics.

In the following, we will summarize briefly the essential equations of the NEGF method. A detailed
introduction can be found, for example, in Refs.~\cite{Datta, Datta2}. The Green's function of the
system is given by
\begin{equation}
  G(E) = (E - H - \Sigma_V - \Sigma_C)^{-1},
\end{equation}   
where $E$ is the energy of the electrons (times a unit matrix), $H$ is the tight-binding Hamiltonian
\eq{Hc} and $\Sigma_V= \sum_n V_n \ket{n}\bra{n}$ the electrostatic potential with 
\begin{equation}
  V_n=
  \begin{cases}
    0 &\text{if  $d_n \leq -w$ (region I)}\\
    \lr{d_n/w+1}V/2 &\text{if  $-w < d_n < w$ (interface)}\\
    V &\text{if $d_n \geq w $ (region II)}
  \end{cases},
\end{equation}
where $d_n$ is the distance to the interface and $2w\sim 5a$ its width. Hence, we assume that on the
atomic scale the junction has some smoothness (see below). In order to suppress boundary effects and
mimic an infinite system, we place a constant complex potential
$\Sigma_C= -\I \sum_{n \in \text{edge}} \ket{n}\bra{n}$ at the edges of the system, which absorbs
the electrons.

\begin{figure*}[t]
  \centering
  \includegraphics[scale= 0.58]{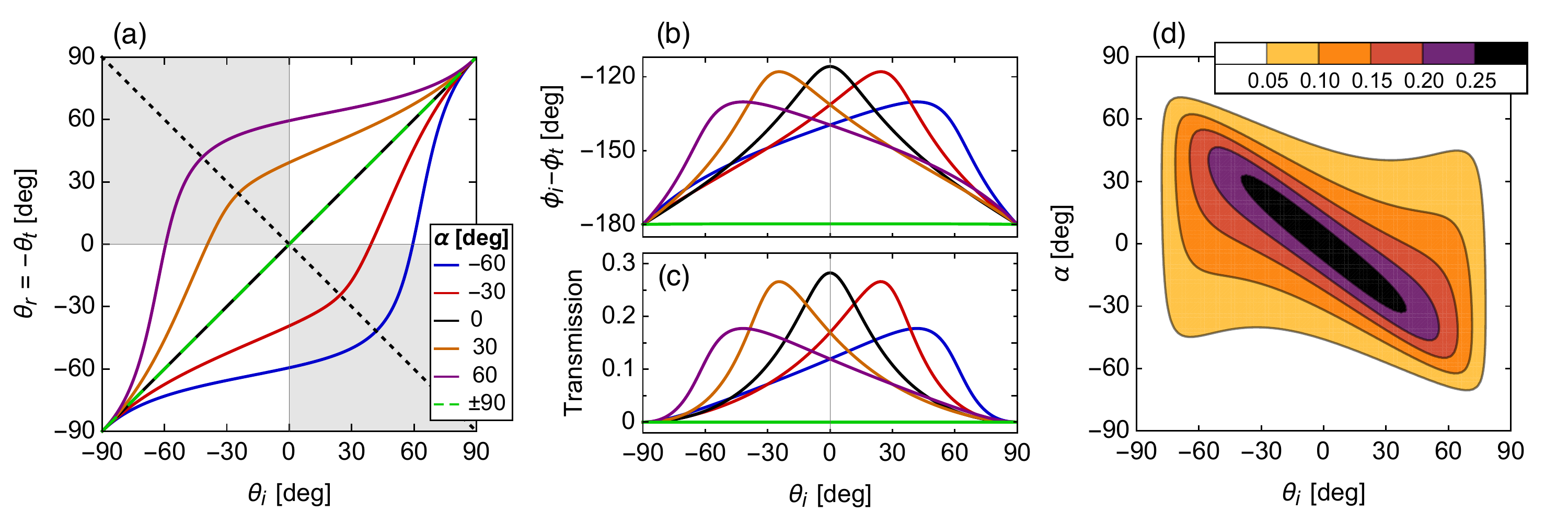}
  \caption{Electron optics laws in a phosphorene $pn$ junction with $V=2E$ and electrons with energy
    $E= 1.2\abs{t_1}$ . (a)~Reflection angle $\th_r$ as a function of the incidence angle $\th_i$
    for various tilting angles $\ap$ of the junction. In general, atypical reflection
    ($\th_r \neq \th_i$) is observed. The curves within the gray shaded squares indicate negative
    reflection. Typical reflection $\th_r= \th_i$ is found only if $\ap= 0\deg, \pm 90\deg$. The
    reflection laws are also valid for asymmetric junctions ($V \neq 2E$), but $\th_t= -\th_r$ holds
    only for the symmetric case. (b)~Difference of the pseudo-spins for the incident and transmitted
    electrons, $\phi_i-\phi_t$. (c,d)~Transmission as a function of $\th_i$ and
    $\ap$. Omnidirectional total reflection, called anti super-Klein tunneling, is observed in
    junctions with $\ap= \pm 90\deg$, because the pseudo-spins of the incident and transmitted
    electrons are anti-parallel. Note that the white regions in (d) correspond to a high degree of
    reflection.}
  \label{fig:4}
\end{figure*}

The electrons are injected at the edges of the system as plane waves propagating towards the
interface of the $pn$ junction. Their momentum $\vec{k}$, which in general does not indicate the
direction of propagation, is calculated from the input parameters $E$ and $\th$ using \eq{ka} and
\eq{ths}. The injection is represented by the inscattering function
\begin{equation}
  \Sigma^{\text{in}}_S = \sum_{n,m \in \text{edge}} A(\vec{r}_n)A(\vec{r}_m) \psi^*_{\vec{k}}(\vec{r}_n)\psi_{\vec{k}}(\vec{r}_m) \ket{n}\bra{m}
\end{equation}
where the sum is over all atoms of the edge where the electrons are injected (for example, all atoms
at the left edge in \fig{3}~(a)). The $\psi_{\vec{k}}(\vec{r}_m)$ are the eigenstates in \eq{wf}
evaluated at the position $\vec{r}_n$ of the atoms at the edge. The function
\begin{equation}
  A(\vec{r}) = \exp\bigl( -\abs{\vec{r} - \vec{r}_{0}}^2/d_0^2 \bigr)
\end{equation}
gives a Gaussian profile to the injected electron wave packet. The parameter $\vec{r}_0$ and $d_0$
control the position and width of the injected electron beam. Finally, the current flowing between
the atoms at positions $\vec{r}_n$ and $\vec{r}_m$ is calculated by means of
\begin{equation}
  I_{nm} = \textrm{Im}(t^*_{nm}G^{\text{in}}_{nm}),
\end{equation}
where
\begin{equation}
  G^{\text{in}} = G \Sigma^{\text{in}}_SG^{\dagger}.
\end{equation}

Good agreement between the quantum current flow and the trajectories of geometric optics can be
expected only in the specific parameter regime, where the Fermi wavelength of the electrons
$\lambda_F$ is much larger than the interatomic distances $a, b$ and the junction width $w$
(justifying the approximations of the continuum and a sharp junction), but smaller than the system
size $\lr{L_x,L_y}$,
\begin{equation}
  a,b,w \: \ll \:  \lambda_F= \frac{2\pi}{\abs{\vec{k}}} \: \ll \: L_x,L_y.
\end{equation}

In the following, we consider in our numerical studies a phosphorene nanoribbon of size
$(L_x,L_y)= 500 \times 500 \,a \approx 220 \times 220 \un{nm}$, which consists of approximately
1.4~million atoms. Electrons are injected at the energy
$E= \Delta+0.2t_1\approx 1.2 \abs{t_1} \approx 1.46\un{eV}$, which corresponds to a Fermi wavelength
$\lambda_F \approx 12a$. The width of the electron beam is $d_0=0.2 L_{x}$. The electron flow in
absence of an electrostatic potential ($V=0$) is shown in \fig{3} and confirms the ballistic,
ray-like propagation of the electrons. Interestingly, we observe that the diffraction of the
electrons is stronger in the $y$ direction (Schr\"odinger fermions) than in the $x$ direction
(massive Dirac fermions). We will address this observation in our future work. From now on, we will
measure all energies in multiples of $\abs{t_1}$ and distances in multiples of $a$.

\section{Negative reflection and anti super-Klein tunneling}
\label{sec:results}

\begin{figure*}[t]
\centering
\includegraphics[scale= 0.46]{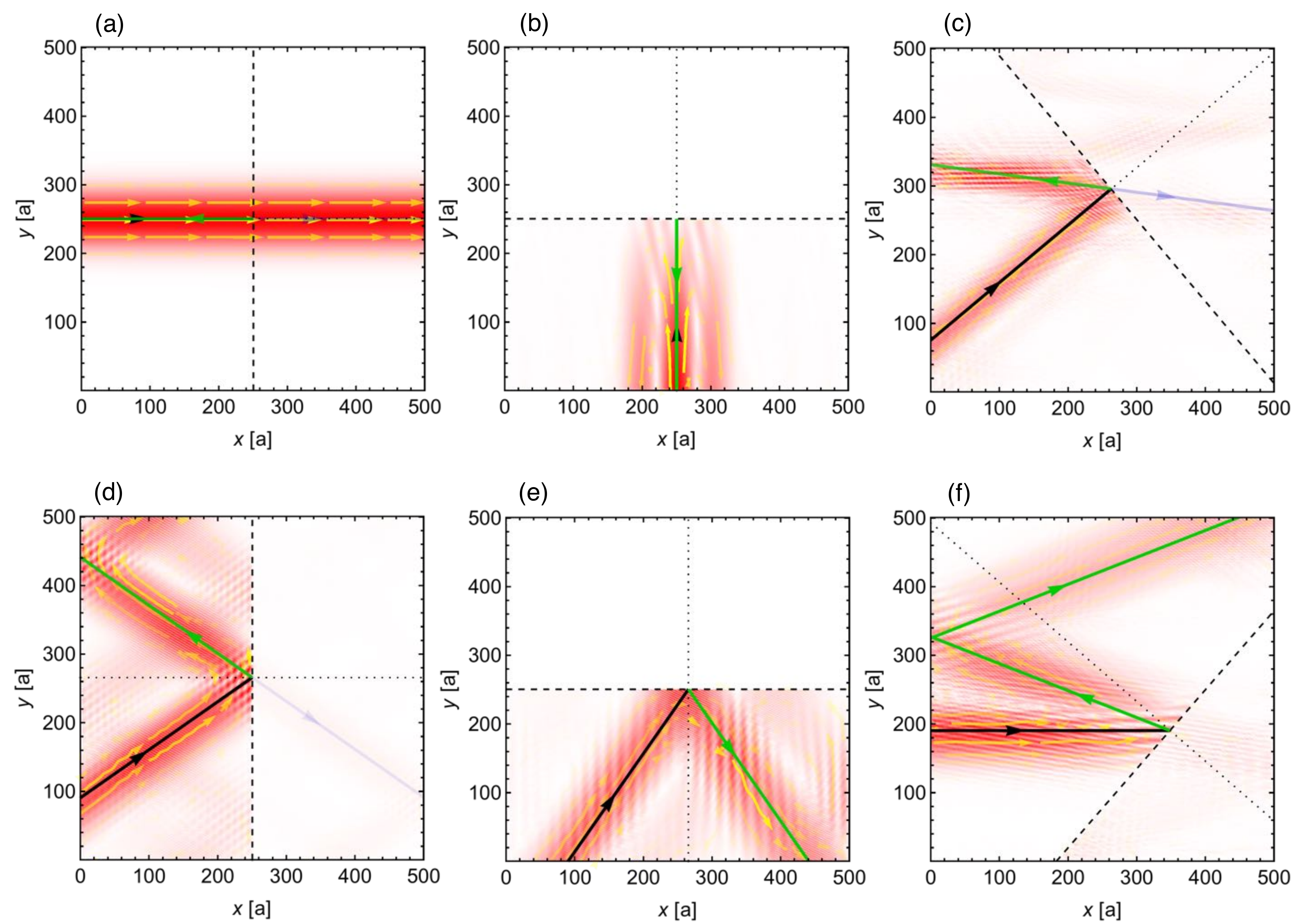}
\caption{Current flow in a phosphorene $pn$ junction with $V=2E$ and $E=1.2\abs{t_1}$ for different
  tiltings of the junction. The interface of the junction is indicated by a black-dashed line and
  its normal by a black-dotted line. The numerically calculated current flow paths (the red color
  shading indicates the current density, the yellow arrows the current vector field), agree
  qualitatively with the ray-like trajectories (solid lines) of geometric optics and hence, confirm
  the electron optics laws derived in Section~\ref{sec:eopt}. (a-c)~Three different cases of normal
  incidence show that the reflection and refraction of the electron beams depend strongly on the
  tilting $\ap$. (d,e)~Typical (specular) reflection is found only if $\ap=0\deg, \pm
  90\deg$. (f)~Negative reflection of the electron beam is observed at the interface of the
  junction, while typical reflection is found at the the edges of the system. (b,e)~For junctions
  with $\ap=\pm 90\deg$, we observe omni-directional total reflection, called anti super-Klein
  tunneling, which is due to the fact that the pseudo-spins of the incident and transmitted
  electrons are anti-parallel.}
\label{fig:5}
\end{figure*}

We evaluate the electron optics laws in phosphorene $pn$ junctions, which were derived in
Section~\ref{sec:eopt}, discuss their properties and compare with the current flow calculated
numerically by the NEGF method. We consider a $pn$ junction with $V=2E$, where the electrons go from
the conduction band in region I to the valence band in region II. The reflection angle $\th_r$ as a
function of the incidence angle $\th_i$ is shown in \fig{4}~(a) for various junction angles
$\ap$. The typical reflection law $ \theta_r = \theta_i $ of optics is found only for junctions
parallel to the $x$ ($\ap= 0\deg$) and $y$ axes ($\ap= \pm 90\deg$), but in general $\th_r$ is a
non-linear function of $\th_i$.

Remarkably, we find negative reflection for certain parameters, see the curves within the gray
shaded regions in \fig{4}~(a). Negative reflection is observed in tilted junctions for incidence
angles $0 < \abs{\th_i} < \abs{\th_M}$, where $\th_M$ is the incidence angle for normal reflection,
see the intersection points of the curves in \fig{4}~(a) with the horizontal $\th_r=0\deg$ line. For
$\abs{\th_i}>\abs{\th_M}$, anomalous reflection is observed, because $\th_i$ and $\th_r$ have the
same sign but different values. Moreover, retroreflection $\th_r= -\th_i$ emerges for certain
incidence angles, see the intersection points of the curves in \fig{4}~(a) with the black-dotted
line. Note that these atypical reflection laws in phosphorene $pn$ junctions are independent of $V$
and hence also valid for junctions with $V \neq 2E$. In symmetric junctions ($V=2E$), we can also
confirm numerically that $\theta_r= -\theta_t$ and hence, negative reflection goes along with
positive refraction and vice versa.

All these unusual properties are confirmed in \fig{5} which shows the current flow paths calculated
numerically by the NEGF method together with the trajectories from geometric optics. The upper
panels (a-c) show three cases of normal incidence but the different orientations of the junction
causes that the reflection and refraction of the electrons are very different. Typical (specular)
reflection is observed in the panels (d,e), while negative reflection is found in panel (f). Note
that the absorption of the electrons at the edges due to the complex potential is not perfect and
hence, typical reflection is observed at the edges. The numerically calculated current densities
show in some cases a ripple pattern, which is on the length scale of the Fermi wavelength and caused
by the interference of electron beams of finite width.

\begin{figure*}[t]
\includegraphics[scale= 0.61]{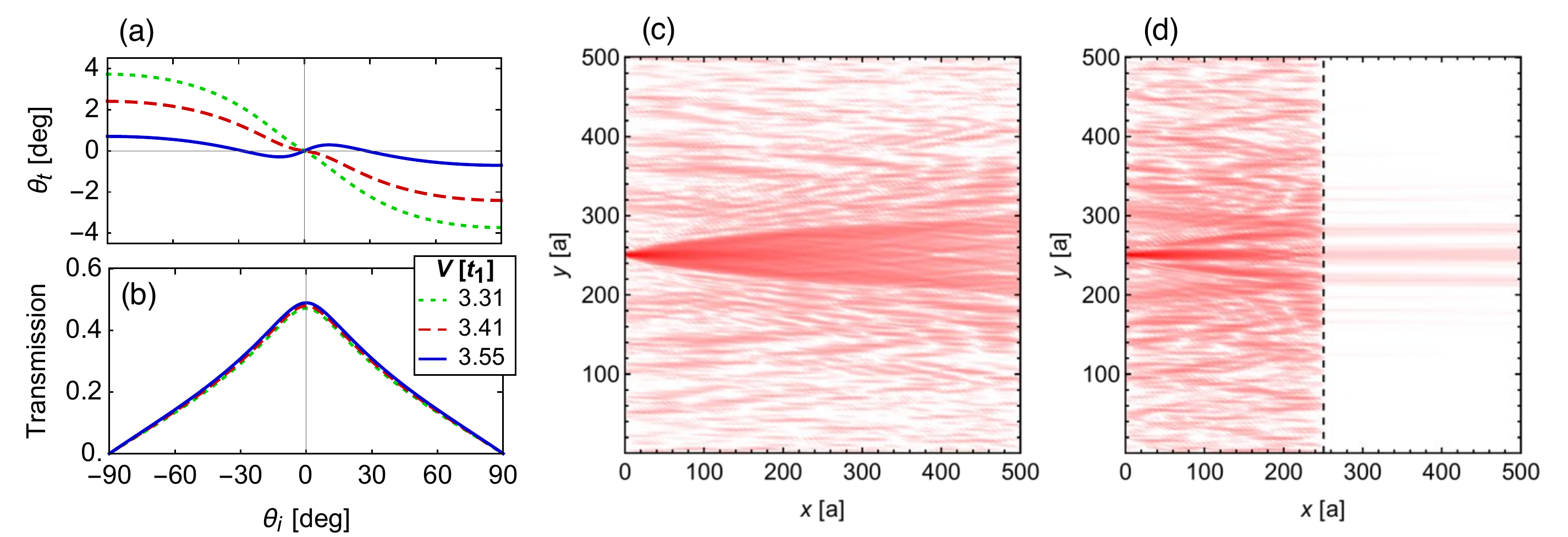}
\caption{Collimation of electrons in phosphorene $pn$ junctions. (a) Refraction angle as a function
  of the incidence angle for electrons with energy $E=1.2\abs{t_1}$ and different values of the
  electrostatic potential $V$. (b) Transmission as a function of the incidence angle. (c,d)
  Numerically calculated current flow for a narrow electron beam in absence (c) and presence (d) of
  the electrostatic potential. Note that a logarithmic scale is used for the red color shading of
  the current density, because the diffraction naturally reduces its density.  All panels show that
  a phosphorene $pn$ junction can act as a collimator, because the electrons are transmitted all in
  the same direction normal to the junction. It also acts as a filter because normally incident
  electrons are transmitted preferentially.}
\label{fig:6}
\end{figure*}

The observed atypical reflection law can be explained by the anisotropy of phosphorene's electronic
structure and the tilting of the junction. Due to these, the intersection points of the blue-dashed
line in \fig{2}~(c) with the constant energy contours represent states having different parallel
group velocity component, although the linear momentum along the interface is conserved. Only if
$\ap=0\deg, \pm 90\deg$, the intersection points represent states with the same parallel group
velocity component causing typical reflection. Note that the atypical reflection can also be
explained by the fact that the tilting of the junction breaks the mirror symmetry with respect to
the $k_x$ and $k_y$ axes in momentum space.

The transmission shown in \fig{4}~(c-d) changes with the tilting of the junction $\ap$. Klein
tunneling, one of the outstanding properties of graphene \cite{Katsnelson, Kim}, is not observed in
phosphorene, because $T<1$. More notably, we observe zero transmission in a junction with
$\ap=\pm 90\deg $ for all incidence angles. This omni-directional total reflection, called
anti-super Klein tunneling, is confirmed numerically in \fig{5}~(b,e). In contrast to the typical
total reflection, it is not caused by the absence of electronic states but can be explained by the
fact that the pseudo-spins of the incident and transmitted electrons are anti-parallel. The absence
of Klein tunneling is also due to these pseudo-spin differences $\phi_i-\phi_t$ shown in
\fig{4}~(b). Interestingly, the difference is quite similar to the transmission, though an exact
scaling law cannot be established. Note that phosphorene's anisotropic energy bands and the tilting
of the $pn$ junction can also cause asymmetric Veselago lenses, such as those predicted for
uniaxially strained graphene \cite{Betancur}.

In asymmetric phosphorene $pn$ junctions, where $V \neq 2E$, further interesting electron-optics
phenomena can be observed. In \fig{6} the electrostatic potential is increased to higher energies
$V \sim 3.5 \abs{t_1}$, while the electron energy $E=1.2 \abs{t_1}$ is kept constant. We observe
that the electrons are refracted in a very narrow window, $\abs{\theta_t} \lesssim 1\deg$, compared to the
broad range of incident angles ranging from $-90 \deg$ to $90 \deg$. The transmission is decaying
almost linearly from $T \approx 0.5$ at normal incidence to zero at $\th_i=\pm 90\deg$, see
\fig{6}~(b). Hence, the $pn$ junction acts a collimator and a filter, because the transmitted
electrons are aligned to flow all in the same direction and the normally incident electrons are
transmitted preferentially. Such collimation effect can be obtained also in a graphene superlattice
\cite{Park} but not in a single graphene $pn$ junction. The numerically calculated current flow in
\fig{6}~(c,d) confirms these effects. A narrow electron beam is injected at the left system edge. In
absence of the electrostatic potential it shows strong diffraction (c), while in presence of the
potential the collimation and filtering of the electrons can be observed clearly (d).

\section{Conclusions}
\label{conclusions}

The ballistic electron flow in phosphorene $pn$ junctions has been studied theoretically. We have
started with a second nearest neighbor tight-binding Hamiltonian that captures the essential
features of phosphorene's electronic structure at low energies, namely a band gap and strong
anisotropy. Due to this anisotropy the electrons behave in one direction as massive Dirac fermions
and in the orthogonal direction as Schr\"odinger electrons, leading to unusual transport properties.

Applying the continuum approximation, we have derived the electron optics laws in phosphorene which
show very particular properties. Because of the anisotropy of the electronic structure, the
orientation of the $pn$ junction with respect to the sublattice is an important parameter, see
\fig{2}. For tilted junctions $0\deg < \abs{\ap} < 90\deg $, we find negative and anomalous
reflection, where $\th_r \neq \th_i$, see \fig{4}. Such atypical properties have been observed for
the reflection of acoustic and electromagnetic waves at the surface of metamaterials
\cite{Liu2,Liu3,Yu,Yu2}, but -- to the best of our knowledge -- not yet for the electron flow in a
nano-material. The typical reflection law $\th_r=\th_i$ is obtained only for junctions aligned with
the $x$ or $y$ axis of the system. Moreover, for junctions parallel to the $y$ axis, we observe
omni-directional total reflection, called anti-super Klein tunneling, which is explained by the fact
that the pseudo-spins are anti-parallel in the two regions of the junction. This effect is the
counterpart of the omni-directional perfect transmission of massless Dirac fermions in Dirac
materials \cite{Betancur3} and pseudo-spin one systems \cite{Bercioux2,Shen,Betancur2,Fang}. We have
also applied the NEGF method on the tight-binding model to calculate numerically the current flow in
finite phosphorene nanoribbons. The good agreement of both approaches confirms clearly the electron
optics laws in phosphorene, see \fig{5}, but also reveals additional interference effects due to the
wave nature of the electron beams.

The exiting transport properties reported in this work will certainly help to pave the way to the
first electron optics experiments in phosphorene $pn$ junctions, but may also have various
nano-technological applications. It has been shown in \fig{6} that the anisotropy of the electronic
structure can be used to construct a collimator and filter of electron beams. The anti-super
Klein tunneling can be used to confine and guide efficiently an electron beam, or to switch its direction by
turning on and off the electrostatic potential.

\vspace*{3mm}
\section*{Acknowledgments}
The authors gratefully acknowledge financial support from CONACYT Proyecto Fronteras 952 and the
UNAM-PAPIIT research grants IN-103017 and IA-101618. We thank Reyes Garcia for computer technical
support and Thomas H. Seligman for helpful discussions.

\bibliography{popt}

\end{document}